\documentclass[prl,twocolumn]{revtex4}
\usepackage[dvips]{graphics}
\usepackage{amsmath}
\usepackage{epic}

\makeatletter
\renewcommand{\@makecaption}[2]{
  \vskip\abovecaptionskip
  \sbox\@tempboxa{\small\sf #1: #2}%
  \ifdim \wd\@tempboxa >\hsize
  \small\sf #1: #2\par
  \else
    \global \@minipagefalse
    \hb@xt@\hsize{\hfil\box\@tempboxa\hfil}%
  \fi
  \vskip\belowcaptionskip}
\makeatother

\newcommand{\tr}{\operatorname{tr}}

\newcommand{\TeV}{\text{ TeV}}

\newsavebox{\moose}
\sbox{\moose}{%
\begin{picture}(0,0)
  \thicklines
  \put(-60,0){\circle{20}}
  \put(60,0){\circle{20}}
  \put(-50,0){\line(1,0){40}}
  \put(0,0){\circle{20}}
  \put(10,0){\line(1,0){40}}
  \put(-25,0){\vector(1,0){0}}
  \put(35,0){\vector(1,0){0}}
\end{picture}}
\newsavebox{\cmoose}
\sbox{\cmoose}{%
\begin{picture}(0,0)
  \thicklines
  \put(-60,0){\circle{30}}
  \put(60,0){\circle{30}}
  \dashline{6}(-45,0)(45,0)
  \put(0,0){\vector(1,0){0}}
\end{picture}}

\begin{document}
\preprint{HUTP-01/A040}
\preprint{BUHEP-01-18}
\preprint{LBNL-48727}

\title{Accelerated Unification}

\author{Nima Arkani-Hamed}
\email{arkani@bose.harvard.edu}
 \thanks{Permanent address: Department of Physics, UC Berkeley, Berkeley,
   CA 94720}
\author{Andrew G. Cohen}%
 \email{cohen@andy.bu.edu}
\thanks{Permanent address: Physics Department, Boston University,
  Boston MA 02215} 
\author{Howard Georgi}
 \email{georgi@physics.harvard.edu}
\affiliation{Lyman  Laboratory of Physics, Harvard University,
  Cambridge MA 02138}

\date{August, 2001}

\begin{abstract}
    We construct four dimensional
    gauge theories in which the successful supersymmetric unification
    of gauge couplings is preserved but accelerated by N-fold
    replication of the MSSM gauge and Higgs  structure. This results
    in a low  unification scale of $10^{13/N} \text{TeV}$.  
\end{abstract}

\maketitle

The unification of gauge couplings~\cite{Georgi:1974sy,Georgi:1974yf}
in the minimal supersymmetric 
standard model (MSSM)~\cite{Dimopoulos:1981zb,Dimopoulos:1981yj} is
one of the strongest 
hints for the structure of physics at and above the TeV scale. This
success is so striking it seems unlikely to have arisen accidentally.
Usually it is argued that this unification depends on the absence of
new particles and interactions beyond those of the MSSM and a desert
stretching from the electroweak scale at $v\sim 1$ TeV to the GUT scale near
$M_G \sim 10^{13}$ TeV.  We have of course never directly observed
unified gauge couplings at such a high scale---instead
we infer this unification by  scaling unified
couplings from 
$M_G$ to the low scale $v$, which  gives the experimentally
successful relations
\begin{align}
  \label{eq:1}
  \frac{2\pi}{\alpha_3(v)}-\frac{2\pi}{\alpha_2(v)} &=
  (b_3-b_2)_{\text{MSSM}} \log(M_G/v) \\
  \label{eq:1a}
  \frac{2\pi}{\alpha_2(v)}-\frac{2\pi}{\alpha_1(v)} &=
  (b_2-b_1)_{\text{MSSM}} \log(M_G/v) 
\end{align}
These relations are obtained by assuming the absence of any charged
matter in the energy desert between $v$ and $M_G$ other than
multiplets, like the MSSM generations, that fill out complete $SU(5)$
representations. Such representations make equal contributions to
$b_{3,2,1}$ and thus affect neither the success of unification nor the
scale $M_G$. Unification thus follows from the MSSM gauge and Higgs
structure. 

Successful unification is usually viewed as strong evidence
for a high fundamental scale.
We show that this conclusion is unwarranted: by replicating the MSSM
gauge and Higgs structure $N$ times we construct models in which the
gauge couplings unify as in the MSSM but at a far lower scale,
$M_U\sim 10^{13/N} \TeV$.

Imagine $N$ copies of the MSSM gauge and Higgs structure, with
fine-structure constants $\alpha_3^{(i)}, \alpha_2^{(i)},
\alpha_1^{(i)}$, where at the scale $M_U$
$\alpha_3^{(i)}=\alpha_2^{(i)}=\alpha_1^{(i)}\equiv \alpha_0^{(i)}$.
The relative running of each copy of the $G_i\equiv SU(3)_i\times
SU(2)_i\times U(1)_i$ gauge couplings is as in the MSSM:
\begin{align}
  \label{eq:2}
  \frac{2\pi}{\alpha_3^{(i)}(\mu)}-\frac{2\pi}{\alpha_2^{(i)}(\mu)} &=
  (b_3-b_2)_{\text{MSSM}} \log(M_U/\mu) \\
  \frac{2\pi}{\alpha_2^{(i)}(\mu)}-\frac{2\pi}{\alpha_1^{(i)}(\mu)} &=
  (b_2-b_1)_{\text{MSSM}} \log(M_U/\mu) 
\end{align}
Now suppose that at the scale $v$ we higgs these $N$
copies of the MSSM gauge group down to the diagonal MSSM gauge group.
The diagonal gauge couplings obtained by tree level matching at the
scale $v$ are:
\begin{equation}
  \label{eq:3}
  \frac{2\pi}{\alpha_{3,2,1}(v)} = \sum_{i=1}^N
  \frac{2\pi}{\alpha_{3,2,1}^{(i)}(v)}
\end{equation}
Combining these equations we obtain the relative size of the low
energy gauge couplings
\begin{align}
  \label{eq:4}
  \frac{2\pi}{\alpha_3(v)}-\frac{2\pi}{\alpha_2(v)} &=
  (b_3-b_2)_{\text{MSSM}} \log \left(M_U^N/v^N\right) \\
  \label{eq:4a}
  \frac{2\pi}{\alpha_2(v)}-\frac{2\pi}{\alpha_1(v)} &=
  (b_2-b_1)_{\text{MSSM}} \log \left(M_U^N/v^N\right)
\end{align}
This  is identical to the MSSM result~\eqref{eq:1},\eqref{eq:1a}
with
\begin{equation}
  \label{eq:5}
  M_U =v  \left(\frac{M_G}{v}\right)^{1/N}\sim 10^{13/N} \TeV  
\end{equation}
In equations~\eqref{eq:4},\eqref{eq:4a} we have
assumed that the fields responsible for higgsing to the
diagonal subgroup fill out complete $SU(5)$ multiplets and do not
affect the relative running.  We have also assumed that the
unification scale is the same for each group $G_i$.

We now describe an explicit model which implements this simple
mechanism.  The model is conveniently summarized in the ``moose'' (or
``quiver'') diagram of figure~\ref{fig:moose}:

\begin{figure}[htbp]
\medskip
  \centering
  \begin{picture}(0,0)
    \thicklines
    \put(-110,0){\circle{25}}
    \put(-110,-25){\makebox(0,0){$G_1$}}
    \put(-50,0){\circle{25}}
    \put(-50,-25){\makebox(0,0){$(G_2)_{X_2}$}}
    \put(10,0){\circle{25}}
    \put(10,-25){\makebox(0,0){$(G_3)_{X_3}$}}
    \put(-100,5){\line(1,0){40}}
    \put(-75,5){\vector(1,0){0}}
    \put(-100,-5){\line(1,0){40}}
    \put(-82,-5){\vector(-1,0){0}}
    \put(-40,5){\line(1,0){40}}
    \put(-15,5){\vector(1,0){0}}
    \put(-40,-5){\line(1,0){40}}
    \put(-22,-5){\vector(-1,0){0}}
    \put(25,0){\makebox(0,0){\dottedline{5}(0,0)(10,0)}}
    \put(50,0){\circle{25}}
    \put(60,5){\line(1,0){40}}
    \put(85,5){\vector(1,0){0}}
    \put(60,-5){\line(1,0){40}}
    \put(78,-5){\vector(-1,0){0}}
    \put(110,0){\circle{25}}
    \put(110,-25){\makebox(0,0){$(G_N)_{X_N}$}}
  \end{picture}
  \bigskip\bigskip
  \caption{A Model}
  \label{fig:moose}
\end{figure}

Each site contains an MSSM gauge group $G_i=SU(3)_i\times
SU(2)_i\times U(1)_i$ and a pair of Higgs doublets. All but the
leftmost site contain an additional $U(1)_{X_i}$. These $U(1)_{X_i}$
gauge groups will be necessary to stabilize our desired pattern of
symmetry breaking.  The charges for the $U(1)_i$ and $U(1)_{X_i}$ will
be chosen orthogonal, eliminating any mixing between the corresponding
gauge bosons.  Since the MSSM generations fill out complete $SU(5)$
multiplets it is not necessary to replicate them. They may be
incorporated into the model in a variety of ways. We might for example
have them all charged under only $G_1$, or we may place them on
different sites in the moose. 

We assume that the couplings of the gauge group $G_i$ unify as before
at a scale $M_U$.  The $U(1)_{X_i}$ coupling may or may not unify with
those of $G_i$ at this scale.  The links with arrows pointing from the
$i$th site to the $(i+1)$th site are chiral superfields $F_i$.
These
fields fill out complete multiplets of the {\em global} $H_i=SU(5)_i$
symmetry into which each $G_i$ is embedded, transforming as
$(\mathbf{5},\mathbf{\bar 5})$ under the global $SU(5)_i \times
SU(5)_{i+1}$.
The
field $F_i$ can be represented by a $5\times 5$ matrix:
\begin{equation}
  \label{eq:6}
  \begin{pmatrix}
    \phi_3  & \phi_x \\
    \phi_y & \phi_2
  \end{pmatrix}
\end{equation}
where under the gauge symmetries $[G_i, G_{i+1}]$ these fields
transform as $\phi_3 \sim \left[(\mathbf{3},1)_{-1/3},(\mathbf{\bar
    3},1)_{1/3}\right]$, $\phi_2 \sim
\left[(1,\mathbf{2})_{1/2},(1,\mathbf{2})_{-1/2}\right]$, $\phi_x \sim
\left[(\mathbf{3},1)_{-1/3}, (1,\mathbf{2})_{-1/2}\right]$, $\phi_y
\sim \left[(1,\mathbf{2})_{1/2},(\mathbf{\bar 3},1)_{1/3}\right]$.
The $F_i$ also have $(U(1)_{X_i},U(1)_{X_{i+1}})$ charges $(1,-1)$.
The links with arrows pointing to the left are chiral superfields
$\bar F_i$ transforming according to the conjugate representation.  As
these fields fill out complete $SU(5)$ multiplets they do not affect
unification.  We may also include additional vector-like fields that
fill out complete multiplets under these global $SU(5)_i$ symmetries.

We include mass terms in the
superpotential for the $F_i, {\bar F}_i$ fields:
\begin{equation}
  \label{eq:7}
  W = \sum_i \tr \mu_i {\bar F}_i F_i
\end{equation}
where for notational convenience we have assumed the masses of all the
components of $F$ are the same. 
In order to higgs down to the MSSM gauge group the fields $\phi_3$ and
$\phi_2$ need to acquire vacuum expectation values $\phi_3^{(i)} =
v_3^{(i)} \mathbf{1}_3$, $\phi_2^{(i)} = v_2^{(i)} \mathbf{1}_2$ with
$v_3^{(i)}, v_2^{(i)}\sim v$.  (Since these fields carry $U(1)$ charges
these vevs suffice to higgs the full MSSM gauge group down to the
diagonal subgroup.)  We can produce these vevs in the same way that a
vev for the MSSM Higgs field is triggered, via soft SUSY breaking
masses ${\tilde m_{3,2,x,y}^2}$.  If $\mu^2 + {\tilde m_{3,2}^2} < 0$
and $\mu^2 + {\tilde m_{x,y}^2} > 0$, the $\phi_{3,2}$ will acquire
vevs. The stabilizing quartic potential is provided by the D-terms.
With only the $SU(3)_i\times SU(2)_i\times U(1)_i$ contributions a
D-flat direction with $v_3^{(i)}=v_2^{(i)}$ would not be
stabilized. The $U(1)_{X_i}$ 
have been included to provide an additional D-term which stabilizes
this direction. 
This higgses the theory down to the diagonal MSSM gauge group.
All the $U(1)_{X_i}$ gauge bosons, together with all the fermionic
components of the $F, {\bar F}$ superfields, become massive.  Thus
this simple theory gives accelerated unification with no exotic states
at very low energies.  

If the  $\mu_i$ are  large compared to the SUSY breaking soft masses,
no vevs are triggered. This is similar to the ``$\mu$'' problem of the
MSSM. We will simply choose the $\mu_i$ to be of order $v$.
Since the vevs which higgs the large gauge symmetry down to the MSSM
gauge group are triggered by SUSY breaking, as is electroweak symmetry
breaking, these vevs are near a TeV. They can be somewhat
larger since they are only stabilized by the $U(1)_X$ D-terms: a small
$U(1)_X$ gauge coupling yields a parametrically larger scale.
If the $U(1)_{X_i}$ gauge couplings also unify with the $G_i$
couplings at the scale $M_U$, they are naturally the smallest
couplings at $v$ since they have the largest $\beta$-function.

At energies somewhat larger than a TeV this theory has many new
particles beyond those of the MSSM, including an $\sim N$-fold spectrum of
massive gauge bosons, Higgs particles, link fields and their
superpartners.

The additional $F_i, {\bar F}_i$ states contribute to the running of
each individual gauge group $G_i$. We should therefore check that all
the gauge couplings remain perturbative at scales between $v$ and
$M_U$. Since the $\alpha_3^{(i)}$ are the largest gauge couplings at
each site at all scales, we need only ensure that these couplings
remain perturbative. For simplicity we will assume that the MSSM
generations are only charged under $G_1$. The $\beta$-function
coefficients for the $\alpha_3^{(i)}$ are $b_3^{(1)} = 2,
b_3^{(2,\dots,N-1)}=1, \text{ and } b_3^{(N)}=-4$. Since all but
$\alpha_3^{(N)}$ are infrared free, it is sufficient to choose the
$\alpha_0^{(i)}$ perturbative for $i=1,\dots,N-1$, 
and to require  $\alpha_3^{(N)}(v)>0$.
Combining with \eqref{eq:3}  and running  up to the scale
$M_U$ gives:
\begin{multline}
  \label{eq:11}
  90 \sim \frac{2\pi}{\alpha_3(v)} +  \log \left(M_G/v\right) \\ > 
  \frac{2\pi}{\alpha_0^{(N)}}  > \frac{4}{N} \log \left(M_G/v\right) \sim
  \frac{120}{N} 
\end{multline}
As long as these inequalities are satisfied the theory remains weakly
coupled between the scales $v$ and $M_U$. These $\beta$-functions and
the condition of perturbativity  preclude the
case where the $\alpha_0^{(i)}$  are all equal. If we had included
additional vector-like multiplets transforming  under
$G_N$, the $\beta$-function coefficient $b_3^{(N)}$ would be different
and could relax these constraints. For example for $b_3^{(N)}=-1$ equal
$\alpha_0^{(i)}$ are possible for any value of $N$.

The general mechanism of accelerated unification is easily
incorporated in other models.  For example an even more minimal model
can be obtained by using link fields that fill out complete multiplets
under the ``trinified'' group $H_i=SU(3)^1_i \times SU(3)^2_i \times
SU(3)^3_i$~\cite{Glashow:1984gc,Lazarides:1993sn} rather than $SU(5)$
multiplets. A simple choice is $F^1_i$ transforming under $[H_i,
H_{i+1}]$ as $[(\mathbf{3},1,1),(\mathbf{\bar 3},1,1)]$ and
$F^{2,3}_i$ defined by cyclic permutation, together with the conjugate
representations ${\bar F^{1,2,3}}_i$.  In addition to mass terms the
superpotential contains cubic interactions $\sum_i \lambda_{1,2,3}
\det F^{1,2,3}_i+\text{conjugate}$.  These renormalizable interactions
lift all the flat directions, eliminating the need for the
$U(1)_{X_i}$ gauge groups of the previous model. SUSY breaking soft
masses which destabilize $F^{1,2}$ produce the correct pattern of
symmetry breaking, higgsing to the diagonal MSSM gauge group with no
exotic particles at low energies. These link fields make smaller
contributions to the $G_i$ $\beta$-functions: for the MSSM generations
all charged under $G_1$, $b_3^{(1)} = 0, b_3^{(2,\dots,N-1)}=-3,
\text{and } b_3^{(N)}=-6$. Therefore all but the first group are
asymptotically free, and all gauge couplings will remain perturbative
at all scales above $v$ as long as
\begin{align}
  \label{eq:12}
  \sum_{i=2}^N \frac{2\pi}{\alpha_0^{(i)}} & < \frac{2\pi}{\alpha_3(v)}
  + 3 \log \left(M_G/v\right) \sim 150 \\
    \label{eq:12a}
  \frac{2\pi}{\alpha_0^{(N)}} &>  \frac{6}{N} \log
    \left(M_G/v\right) \sim \frac{180}{N} \\ 
    \label{eq:12b}
  \frac{2\pi}{\alpha_0^{(2,\dots,N-1)}} &>  \frac{3}{N} \log
    \left(M_G/v\right) \sim \frac{90}{N}
\end{align}
Again the case of equal $\alpha_0^{(i)}$ is precluded without the
addition of extra fields charged under $G_N$.

The formulas \eqref{eq:4},\eqref{eq:4a} do not include various threshold
corrections~\cite{Hall:1981kf}. Although these corrections depend on
the details of the model such as the complete mass spectrum, the
general size of these effects are easily estimated. The unification
relation is most conveniently expressed as a prediction for $\alpha_3$
in terms of the more accurately measured couplings $\alpha_{1,2}$. The
threshold corrections are similar in size to those of the MSSM, except
multiplied by $N$. The resulting prediction for $\alpha_3$ is
\begin{equation}
  \label{eq:9}
  \frac{2\pi}{\alpha_3} = \frac{b_3 -b_2}{b_2-b_1}
  \left(\frac{2\pi}{\alpha_2}- \frac{2\pi}{\alpha_1}\right) +
  \frac{2\pi}{\alpha_2} + C N 
\end{equation}
where the last term $C N$ represents the threshold corrections and $C$
is a constant which depends on group theory factors and the detailed
mass spectrum of the theory, but is parametrically independent of $N$.
In addition to these threshold corrections there are two-loop running
corrections of comparable magnitude.  
These  terms produce a fractional change in $\alpha_3$ of
size
\begin{equation}
  \label{eq:10}
  \frac{\delta \alpha_3}{\alpha_3} = - C N \frac{\alpha_3}{2\pi}
  \simeq -\frac{C N}{60}.
\end{equation}
Roughly speaking, since the size of the matter content under each
gauge group is not dramatically different from the MSSM, these
corrections are of order $N$ times larger than the MSSM threshold
corrections.  Note that this means we cannot take $N$ arbitrarily
large---if $N$ becomes too large these threshold corrections would
destroy the successful MSSM prediction unless the parameters conspire
to make $C$ small, in which case the success of the MSSM would be an
accident.  The precise size of $C$ and the corresponding limit on $N$
is model-dependent.  For $N$ as small as 2, these corrections should
be negligible for most reasonable choices of parameters, while for $N$
as large as 60 the opposite is true.  Intermediate values interpolate
between these two extremes.  The exponential dependence of $M_U$ on
$N$ means that a low unification scale does not require a large
value of $N$: already for $N$ of 6, the unification scale is near
$100$ TeV.

Given the low unification scale~\eqref{eq:5} we might wonder about
other unification phenomena. A qualitative success of the usual desert
is the proximity of the GUT scale and the Planck scale, a relation
absent here. It is nevertheless possible that the Planck/string scale
is $M_U$, with the usual phenomenology associated with a low quantum
gravity scale~\cite{Arkani-Hamed:1998rs}.  With such a low scale,
proton decay becomes an important constraint.  This can be dealt with
in a number of ways familiar from theories of TeV scale quantum
gravity (see for example~\cite{Arkani-Hamed:1999dc}).

It is also possible that the quantum gravity scale remains high, with
a purely gauge theoretic unification at $M_U$. Proton decay can be
avoided in this case as well.  We might consider unifying all but the
left-most group in figure~\ref{fig:moose} into $SU(5)$.  By placing
the MSSM generations on this first site we can enforce baryon number
symmetries which prevent dangerous proton decay~\cite{Weiner:2001pv}.
Or we could contemplate a model in which each $G_i$ is unified into a
trinified group.

Accelerated unification with a high fundamental scale for
4-dimensional quantum gravity allows for new physics between $M_U$ and
$M_{\text{Planck}}$. Unlike conventional unification which requires a
desert from 1 TeV to $M_G$, our models allow even strongly coupled
dynamics at intermediate scales. An example might be the strongly
coupled CFT dynamics  dual to RS type models~\cite{Randall:1999ee}.

The model of figure~\ref{fig:moose} looks very much like a
deconstructed extra dimension~\cite{Arkani-Hamed:2001ca,Hill:2000mu}.
Extra-dimensional models with a low unification scale have been
proposed in~\cite{Dienes:1998vh}, where the power-law running of the
gauge couplings in a (compactified) extra dimension provides faster
unification. In straightforward deconstructions of these power-law
unification models, holomorphy of the resulting 4-d supersymmetric
moose model allows reliable computation of the running of the gauge
couplings, even at strong coupling. The results confirm the power-law
running of~\cite{Dienes:1998vh}, but demonstrate UV sensitivity to the
precise way in which the 5-dimensional theory is
completed~\cite{comm}.  Power-law running does occur in the regime
where the theory looks 5-dimensional. However in deconstructed
examples, corrections from the scale where the theory transitions to a
4-dimensional gauge theory typically destroy the unification of the
couplings.  For example in two different deconstructions, which look
identical below the scale at which the 5th dimension forms, the values
of the high energy MSSM gauge couplings are completely different.

By contrast in our model of accelerated running, the phenomena which
results in a low unification scale is occurring at energies {\em
  above} the scale where, for large $N$, a 5th dimension would form.
The power-law running effects from below this scale are incorporated
in the threshold corrections proportional to $N$ in
equation~\eqref{eq:9}.  For small $N$ the accelerated unification
effects, which come from running at high energies, dominate over these
threshold effects.

Although inspired by deconstruction of power-law running in higher
dimensional field theories, accelerated unification demonstrates a
completely non-extra-dimensional phenomenon.  As emphasized
in~\cite{Arkani-Hamed:2001ca}, ``theory space'' provides a much richer set
of possibilities than field-theoretic extra dimensions, which arise as
a special case. As in other
examples~\cite{Arkani-Hamed:2001nc,Cheng:2001nh,Csaki:2001em,Cheng:2001an,Csaki:2001qm,Cheng:2001qp}
the power of deconstruction lies in its ability to construct purely
four-dimensional models exhibiting surprising field-theoretic phenomena,
with no higher-dimensional interpretation.

The phenomenology of accelerated unification models at the TeV scale
involves a plethora of new particles beyond those of the MSSM.  It is
these particles at the TeV scale which are responsible for lowering
the unification scale to more accessible energies. Exploration of this
phenomenology offers exciting opportunities for future experiments.


\bigskip We thank Ann Nelson for useful discussions.  We also thank
Lisa Randall for informing us of recent work on 
unification with an extra dimension.  H.G. is supported in part
by the National Science Foundation under grant number
NSF-PHY/98-02709. A.G.C. is supported in part by the Department of
Energy under grant number \#DE-FG02-91ER-40676.  N.A-H.  is supported
in part by the Department of Energy. under Contracts
DE-AC03-76SF00098, the National Science Foundation under grant
PHY-95-14797, the Alfred P. Sloan foundation, and the David and
Lucille Packard Foundation.


\begin{thebibliography}{21}
\expandafter\ifx\csname natexlab\endcsname\relax\def\natexlab#1{#1}\fi
\expandafter\ifx\csname bibnamefont\endcsname\relax
  \def\bibnamefont#1{#1}\fi
\expandafter\ifx\csname bibfnamefont\endcsname\relax
  \def\bibfnamefont#1{#1}\fi
\expandafter\ifx\csname citenamefont\endcsname\relax
  \def\citenamefont#1{#1}\fi
\expandafter\ifx\csname url\endcsname\relax
  \def\url#1{\texttt{#1}}\fi
\expandafter\ifx\csname urlprefix\endcsname\relax\def\urlprefix{URL }\fi
\providecommand{\bibinfo}[2]{#2}
\providecommand{\eprint}[2][]{\url{#2}}

\bibitem[{\citenamefont{Georgi and Glashow}(1974)}]{Georgi:1974sy}
\bibinfo{author}{\bibfnamefont{H.}~\bibnamefont{Georgi}} \bibnamefont{and}
  \bibinfo{author}{\bibfnamefont{S.~L.} \bibnamefont{Glashow}},
  \bibinfo{journal}{Phys. Rev. Lett.} \textbf{\bibinfo{volume}{32}},
  \bibinfo{pages}{438} (\bibinfo{year}{1974}).

\bibitem[{\citenamefont{Georgi et~al.}(1974)\citenamefont{Georgi, Quinn, and
  Weinberg}}]{Georgi:1974yf}
\bibinfo{author}{\bibfnamefont{H.}~\bibnamefont{Georgi}},
  \bibinfo{author}{\bibfnamefont{H.~R.} \bibnamefont{Quinn}}, \bibnamefont{and}
  \bibinfo{author}{\bibfnamefont{S.}~\bibnamefont{Weinberg}},
  \bibinfo{journal}{Phys. Rev. Lett.} \textbf{\bibinfo{volume}{33}},
  \bibinfo{pages}{451} (\bibinfo{year}{1974}).

\bibitem[{\citenamefont{Dimopoulos and Georgi}(1981)}]{Dimopoulos:1981zb}
\bibinfo{author}{\bibfnamefont{S.}~\bibnamefont{Dimopoulos}} \bibnamefont{and}
  \bibinfo{author}{\bibfnamefont{H.}~\bibnamefont{Georgi}},
  \bibinfo{journal}{Nucl. Phys.} \textbf{\bibinfo{volume}{B193}},
  \bibinfo{pages}{150} (\bibinfo{year}{1981}).

\bibitem[{\citenamefont{Dimopoulos et~al.}(1981)\citenamefont{Dimopoulos, Raby,
  and Wilczek}}]{Dimopoulos:1981yj}
\bibinfo{author}{\bibfnamefont{S.}~\bibnamefont{Dimopoulos}},
  \bibinfo{author}{\bibfnamefont{S.}~\bibnamefont{Raby}}, \bibnamefont{and}
  \bibinfo{author}{\bibfnamefont{F.}~\bibnamefont{Wilczek}},
  \bibinfo{journal}{Phys. Rev.} \textbf{\bibinfo{volume}{D24}},
  \bibinfo{pages}{1681} (\bibinfo{year}{1981}).

\bibitem[{\citenamefont{Glashow}(1984)}]{Glashow:1984gc}
\bibinfo{author}{\bibfnamefont{S.~L.} \bibnamefont{Glashow}}, in
  \emph{\bibinfo{booktitle}{FIFTH WORKSHOP ON GRAND UNIFICATION}}, edited by
  \bibinfo{editor}{\bibfnamefont{K.}~\bibnamefont{Kang}},
  \bibinfo{editor}{\bibfnamefont{H.}~\bibnamefont{Fried}}, \bibnamefont{and}
  \bibinfo{editor}{\bibfnamefont{P.}~\bibnamefont{Frampton}}
  (\bibinfo{publisher}{World Scientific}, \bibinfo{year}{1984}).

\bibitem[{\citenamefont{Lazarides et~al.}(1993)\citenamefont{Lazarides,
  Panagiotakopoulos, and Shafi}}]{Lazarides:1993sn}
\bibinfo{author}{\bibfnamefont{G.}~\bibnamefont{Lazarides}},
  \bibinfo{author}{\bibfnamefont{C.}~\bibnamefont{Panagiotakopoulos}},
  \bibnamefont{and} \bibinfo{author}{\bibfnamefont{Q.}~\bibnamefont{Shafi}},
  \bibinfo{journal}{Phys. Lett.} \textbf{\bibinfo{volume}{B315}},
  \bibinfo{pages}{325} (\bibinfo{year}{1993}), \eprint{hep-ph/9306332}.

\bibitem[{\citenamefont{Hall}(1981)}]{Hall:1981kf}
\bibinfo{author}{\bibfnamefont{L.}~\bibnamefont{Hall}}, \bibinfo{journal}{Nucl.
  Phys.} \textbf{\bibinfo{volume}{B178}}, \bibinfo{pages}{75}
  (\bibinfo{year}{1981}).

\bibitem[{\citenamefont{Arkani-Hamed et~al.}(1998)\citenamefont{Arkani-Hamed,
  Dimopoulos, and Dvali}}]{Arkani-Hamed:1998rs}
\bibinfo{author}{\bibfnamefont{N.}~\bibnamefont{Arkani-Hamed}},
  \bibinfo{author}{\bibfnamefont{S.}~\bibnamefont{Dimopoulos}},
  \bibnamefont{and} \bibinfo{author}{\bibfnamefont{G.}~\bibnamefont{Dvali}},
  \bibinfo{journal}{Phys. Lett.} \textbf{\bibinfo{volume}{B429}},
  \bibinfo{pages}{263} (\bibinfo{year}{1998}), \eprint{hep-ph/9803315}.

\bibitem[{\citenamefont{Arkani-Hamed and Schmaltz}(2000)}]{Arkani-Hamed:1999dc}
\bibinfo{author}{\bibfnamefont{N.}~\bibnamefont{Arkani-Hamed}}
  \bibnamefont{and} \bibinfo{author}{\bibfnamefont{M.}~\bibnamefont{Schmaltz}},
  \bibinfo{journal}{Phys. Rev.} \textbf{\bibinfo{volume}{D61}},
  \bibinfo{pages}{033005} (\bibinfo{year}{2000}), \eprint{hep-ph/9903417}.

\bibitem[{\citenamefont{Weiner}(2001)}]{Weiner:2001pv}
\bibinfo{author}{\bibfnamefont{N.}~\bibnamefont{Weiner}}
  (\bibinfo{year}{2001}), \eprint{hep-ph/0106097}.

\bibitem[{\citenamefont{Randall and Sundrum}(1999)}]{Randall:1999ee}
\bibinfo{author}{\bibfnamefont{L.}~\bibnamefont{Randall}} \bibnamefont{and}
  \bibinfo{author}{\bibfnamefont{R.}~\bibnamefont{Sundrum}},
  \bibinfo{journal}{Phys. Rev. Lett.} \textbf{\bibinfo{volume}{83}},
  \bibinfo{pages}{3370} (\bibinfo{year}{1999}), \eprint{hep-ph/9905221}.

\bibitem[{\citenamefont{Arkani-Hamed
  et~al.}(2001{\natexlab{a}})\citenamefont{Arkani-Hamed, Cohen, and
  Georgi}}]{Arkani-Hamed:2001ca}
\bibinfo{author}{\bibfnamefont{N.}~\bibnamefont{Arkani-Hamed}},
  \bibinfo{author}{\bibfnamefont{A.~G.} \bibnamefont{Cohen}}, \bibnamefont{and}
  \bibinfo{author}{\bibfnamefont{H.}~\bibnamefont{Georgi}},
  \bibinfo{journal}{Phys. Rev. Lett.} \textbf{\bibinfo{volume}{86}},
  \bibinfo{pages}{4757} (\bibinfo{year}{2001}{\natexlab{a}}),
  \eprint{hep-th/0104005}.

\bibitem[{\citenamefont{Hill et~al.}(2001)\citenamefont{Hill, Pokorski, and
  Wang}}]{Hill:2000mu}
\bibinfo{author}{\bibfnamefont{C.~T.} \bibnamefont{Hill}},
  \bibinfo{author}{\bibfnamefont{S.}~\bibnamefont{Pokorski}}, \bibnamefont{and}
  \bibinfo{author}{\bibfnamefont{J.}~\bibnamefont{Wang}}
  (\bibinfo{year}{2001}), \eprint{hep-th/0104035}.

\bibitem[{\citenamefont{Dienes et~al.}(1998)\citenamefont{Dienes, Dudas, and
  Gherghetta}}]{Dienes:1998vh}
\bibinfo{author}{\bibfnamefont{K.~R.} \bibnamefont{Dienes}},
  \bibinfo{author}{\bibfnamefont{E.}~\bibnamefont{Dudas}}, \bibnamefont{and}
  \bibinfo{author}{\bibfnamefont{T.}~\bibnamefont{Gherghetta}},
  \bibinfo{journal}{Phys. Lett.} \textbf{\bibinfo{volume}{B436}},
  \bibinfo{pages}{55} (\bibinfo{year}{1998}), \eprint{hep-ph/9803466}.

\bibitem[{\citenamefont{Arkani-Hamed
  et~al.}(2001{\natexlab{b}})\citenamefont{Arkani-Hamed, Cohen, and
  Georgi}}]{comm}
\bibinfo{author}{\bibfnamefont{N.}~\bibnamefont{Arkani-Hamed}},
  \bibinfo{author}{\bibfnamefont{A.~G.} \bibnamefont{Cohen}}, \bibnamefont{and}
  \bibinfo{author}{\bibfnamefont{H.}~\bibnamefont{Georgi}},
  \emph{\bibinfo{title}{Private communication}}
  (\bibinfo{year}{2001}{\natexlab{b}}).

\bibitem[{\citenamefont{Arkani-Hamed
  et~al.}(2001{\natexlab{c}})\citenamefont{Arkani-Hamed, Cohen, and
  Georgi}}]{Arkani-Hamed:2001nc}
\bibinfo{author}{\bibfnamefont{N.}~\bibnamefont{Arkani-Hamed}},
  \bibinfo{author}{\bibfnamefont{A.~G.} \bibnamefont{Cohen}}, \bibnamefont{and}
  \bibinfo{author}{\bibfnamefont{H.}~\bibnamefont{Georgi}},
  \bibinfo{journal}{Phys. Lett.} \textbf{\bibinfo{volume}{B513}},
  \bibinfo{pages}{232} (\bibinfo{year}{2001}{\natexlab{c}}),
  \eprint{hep-ph/0105239}.

\bibitem[{\citenamefont{Cheng et~al.}(2001{\natexlab{a}})\citenamefont{Cheng,
  Hill, and Wang}}]{Cheng:2001nh}
\bibinfo{author}{\bibfnamefont{H.-C.} \bibnamefont{Cheng}},
  \bibinfo{author}{\bibfnamefont{C.~T.} \bibnamefont{Hill}}, \bibnamefont{and}
  \bibinfo{author}{\bibfnamefont{J.}~\bibnamefont{Wang}}
  (\bibinfo{year}{2001}{\natexlab{a}}), \eprint{hep-ph/0105323}.

\bibitem[{\citenamefont{Csaki et~al.}(2001{\natexlab{a}})\citenamefont{Csaki,
  Erlich, Grojean, and Kribs}}]{Csaki:2001em}
\bibinfo{author}{\bibfnamefont{C.}~\bibnamefont{Csaki}},
  \bibinfo{author}{\bibfnamefont{J.}~\bibnamefont{Erlich}},
  \bibinfo{author}{\bibfnamefont{C.}~\bibnamefont{Grojean}}, \bibnamefont{and}
  \bibinfo{author}{\bibfnamefont{G.~D.} \bibnamefont{Kribs}}
  (\bibinfo{year}{2001}{\natexlab{a}}), \eprint{hep-ph/0106044}.

\bibitem[{\citenamefont{Cheng et~al.}(2001{\natexlab{b}})\citenamefont{Cheng,
  Kaplan, Schmaltz, and Skiba}}]{Cheng:2001an}
\bibinfo{author}{\bibfnamefont{H.~C.} \bibnamefont{Cheng}},
  \bibinfo{author}{\bibfnamefont{D.~E.} \bibnamefont{Kaplan}},
  \bibinfo{author}{\bibfnamefont{M.}~\bibnamefont{Schmaltz}}, \bibnamefont{and}
  \bibinfo{author}{\bibfnamefont{W.}~\bibnamefont{Skiba}}
  (\bibinfo{year}{2001}{\natexlab{b}}), \eprint{hep-ph/0106098}.

\bibitem[{\citenamefont{Csaki et~al.}(2001{\natexlab{b}})\citenamefont{Csaki,
  Kribs, and Terning}}]{Csaki:2001qm}
\bibinfo{author}{\bibfnamefont{C.}~\bibnamefont{Csaki}},
  \bibinfo{author}{\bibfnamefont{G.~D.} \bibnamefont{Kribs}}, \bibnamefont{and}
  \bibinfo{author}{\bibfnamefont{J.}~\bibnamefont{Terning}}
  (\bibinfo{year}{2001}{\natexlab{b}}), \eprint{hep-ph/0107266}.

\bibitem[{\citenamefont{Cheng et~al.}(2001{\natexlab{c}})\citenamefont{Cheng,
  Matchev, and Wang}}]{Cheng:2001qp}
\bibinfo{author}{\bibfnamefont{H.-C.} \bibnamefont{Cheng}},
  \bibinfo{author}{\bibfnamefont{K.~T.} \bibnamefont{Matchev}},
  \bibnamefont{and} \bibinfo{author}{\bibfnamefont{J.}~\bibnamefont{Wang}}
  (\bibinfo{year}{2001}{\natexlab{c}}), \eprint{hep-ph/0107268}.

\end{thebibliography}

\end{document}